\documentclass[aps,prl,twocolumn,showpacs,amsmath,amssymb,titlepage]{revtex4}
\usepackage{graphicx}
\usepackage{dcolumn}
\usepackage{bm}

\newcommand{\PT}{\mathcal{PT}}

\begin{document}

\title{$\PT$-symmetric photonic quantum systems with gain and loss\\ do not exist}

\author{Stefan Scheel}
\author{Alexander Szameit}
\address{Institut f\"ur Physik, Universit\"at Rostock,
Albert-Einstein-Stra{\ss}e 23, D-18059 Rostock, Germany \vspace{1cm}}

\begin{abstract}
We discuss the impact of gain and loss on the evolution of photonic quantum states and find that $\PT$-symmetric quantum optics in gain/loss systems is not possible. Within the framework of macroscopic quantum electrodynamics we show that gain and loss are associated with non-compact and compact operator transformations, respectively. This implies a fundamentally different way in which quantum correlations between a quantum system and a reservoir are built up and destroyed.
\end{abstract}


\maketitle

In 1998, Carl Bender challenged the perceived wisdom of quantum mechanics that the Hamiltonian operator describing any quantum mechanical system has to be Hermitian \cite{Bender}. He showed that, in order to possess a real eigenvalue spectrum, a Hamiltonian does not have to be
Hermitian: also Hamiltonians that are invariant under combined parity-time
($\PT$) symmetry operations have this property \cite{Bender02,Bender07}. These findings had
profound impact in particular on photonics research where the required
potential landscapes can be easily generated by appropriately distributing gain
and loss for electromagnetic waves. On this ground it was possible to show,
for example, the existence of non-orthogonal eigenmodes \cite{Kip},
non-reciprocal light evolution \cite{Regensburger2012}, and $\PT$-symmetric lasers
\cite{Feng14,Hodaei14}. Even in fields beyond photonics $\PT$-symmetry has an impact, ranging from
$\PT$-symmetric atomic diffusion \cite{Zhao07}, superconducting wires
\cite{Rubenstein07,Chtchelkatchev12}, and $\PT$-symmetric electronic circuits
\cite{Schindler11}. However, all these are classical phenomena as only single
electromagnetic wave packets are involved. The experimental demonstration of
truly quantum features in $\PT$-symmetric systems with gain and loss is still elusive. Here we
show that this will remain to be the case. Our investigations unequivocally
prove that the common approach for realising $\PT$-symmetric systems in
photonics by concatenating lossy and amplifying media always results in
thermally broadened quantum states. $\PT$-symmetric quantum optics in gain/loss systems is
therefore not possible.

$\PT$-symmetric systems are described by a Hamiltonian that is invariant under
parity-time symmetry transformations \cite{Bender}. In a more mathematical language
this means that if the Hamiltionian $\hat{H}$ commutes with the $\PT$-operator: $\lbrack \hat{H},\hat{\PT}
\rbrack = 0$ and the Hamiltonian and the $\PT$-operator share the same set of
eigenstates, then the eigenvalues of $\hat{H}$ are entirely real. A necessary condition for this symmetry to hold is that the
underlying potential obeys the relation $\hat{V}(-x)=\hat{V}^\ast(x)$. A commonly used example is the complex
anharmonic potential $\hat{V}(x)=ix^3$. $\PT$-symmetry is perceived as a
complex extension of Hermitian quantum mechanics as it also provides a unitary
time evolution \cite{Bender02}. Whereas complex potentials are difficult to
realise in most physical systems, in 2007 it was shown that photonics provides
a suitable testing ground due to the complex nature of the refractive index
\cite{ElGanainy07,Makris08}. Since then, $\PT$-symmetric systems have been
explored in a variety of photonics platforms, ranging from waveguide arrays
\cite{Kip}, fiber lattices \cite{Regensburger2012}, coupled optical resonators
\cite{Peng14}, plasmonics \cite{Benisty11} and microwave cavities
\cite{Bittner12}. Besides the broad range of platforms and observed phenomena,
all these systems make use of classical electromagnetic waves. It it still an
open question as to whether or not quantised light shows the same behaviour,
although recent works imply that it might not as systems with a $\PT$-symmetric Hamiltonian were shown to emit radiation \cite{Schomerus10,Agarwal12}.
What we will unequivocally show is that indeed in all these platforms quantum optical $\PT$-symmetry
does not exist.

The implementation of $\PT$-symmetry in photonics is based on the observation
that the Schr\"odinger equation of quantum mechanics and the Helmholtz equation
of electromagnetism are formally equivalent if the potential $\hat{V}(x)$ in
the Schr\"odinger equation is replaced by the refractive index profile $n(x)$
in the Helmholtz equation \cite{Schrodinger26}. $\PT$-symmetry then translates
into the condition for the complex refractive index $n(-x)=n^\ast(x)$, in
particular, the real part $n_R(x)$ is symmetric and the imaginary part $n_I(x)$
is antisymmetric under the parity operation. The latter implies that loss in
one propagation direction has to be compensated by an identical gain in the
opposite direction \cite{ElGanainy07}. Whereas this concept is well-defined for
the amplitudes of classical electromagnetic waves, this is no longer the case
for the amplitude operators of quantum states of light as they have to obey
certain commutation relations. For example, the amplitude operators for a
single harmonic oscillator have to fulfil the relation
$[\hat{a},\hat{a}^\dagger]=1$ for all times. However, phenomenologically a
dissipation process is always accompanied by additional (Langevin) noise. Hence,
the evolution equation for a damped harmonic oscillator mode with frequency
$\omega$ has to be written as
\begin{equation}
\label{eq:langevin}
\dot{\hat{a}} = (-i\omega-\Gamma) \hat{a} +\hat{f}
\end{equation}
with properly chosen commutation relations between the harmonic oscillator
mode and the noise operators, and where the fluctuation strength of the noise
operator $\hat{f}$ is related to the damping rate $\Gamma$ \cite{Breuer}.

The appropriate framework in which to describe the propagation of quantum
states of light through absorbing and amplifying media is macroscopic quantum
electrodynamics \cite{Scheel98,Acta}. Here the creation and annihilation
operators of the free electromagnetic field have to be replaced by new operators
that describe the collective excitation of the field and the absorbing or
amplifying matter. Within the framework of linear response this theory is exact.
The result is a proper identification of the parameters $\Gamma$ and $\hat{f}$
in Eq.~(\ref{eq:langevin}) by phenomenological quantities such as absorption
and transmission coefficients. This theory provides the basis for the
propagation of quantum states of light through absorbing and amplifying media
\cite{Knoll99,Scheel00}. One first constructs a unitary operation in a larger
Hilbert space of field and medium operators which, after projecting onto the
field quantities alone, results in an effective, typically non-unitary
evolution of the quantum states of light \cite{Knoll99}.
Although the formalism is very similar for absorbing and amplifying media,
there are crucial differences between them that impact the $\PT$-symmetry.
Viewing an optical element as a four-port device with two input and two output
channels for light of a given frequency $\omega$ (note that in a linearly
responding medium light modes of different frequencies do not mix), the
quantum-state transformation at absorbing media corresponds to a compact SU(4)
transformation \cite{Knoll99} whereas the equivalent relation at amplifying
media is a non-compact SU(2,2) transformation \cite{Scheel00}. This seemingly
inoccuous difference has far-reaching consequences: an initial coherent
quantum state $|a_0\rangle$, after propagation through an absorbing medium
with transmission coefficient $T$, remains a coherent quantum state, albeit
with diminished coherent amplitude $|Ta_0\rangle$. On the contrary, after
propagation through an amplifying medium, a coherent state turns into a
displaced thermal state with an effective temperature that depends on the gain
(for details of the calculation, see Supplementary Material).

\begin{figure}[h!]
\includegraphics[width=\columnwidth]{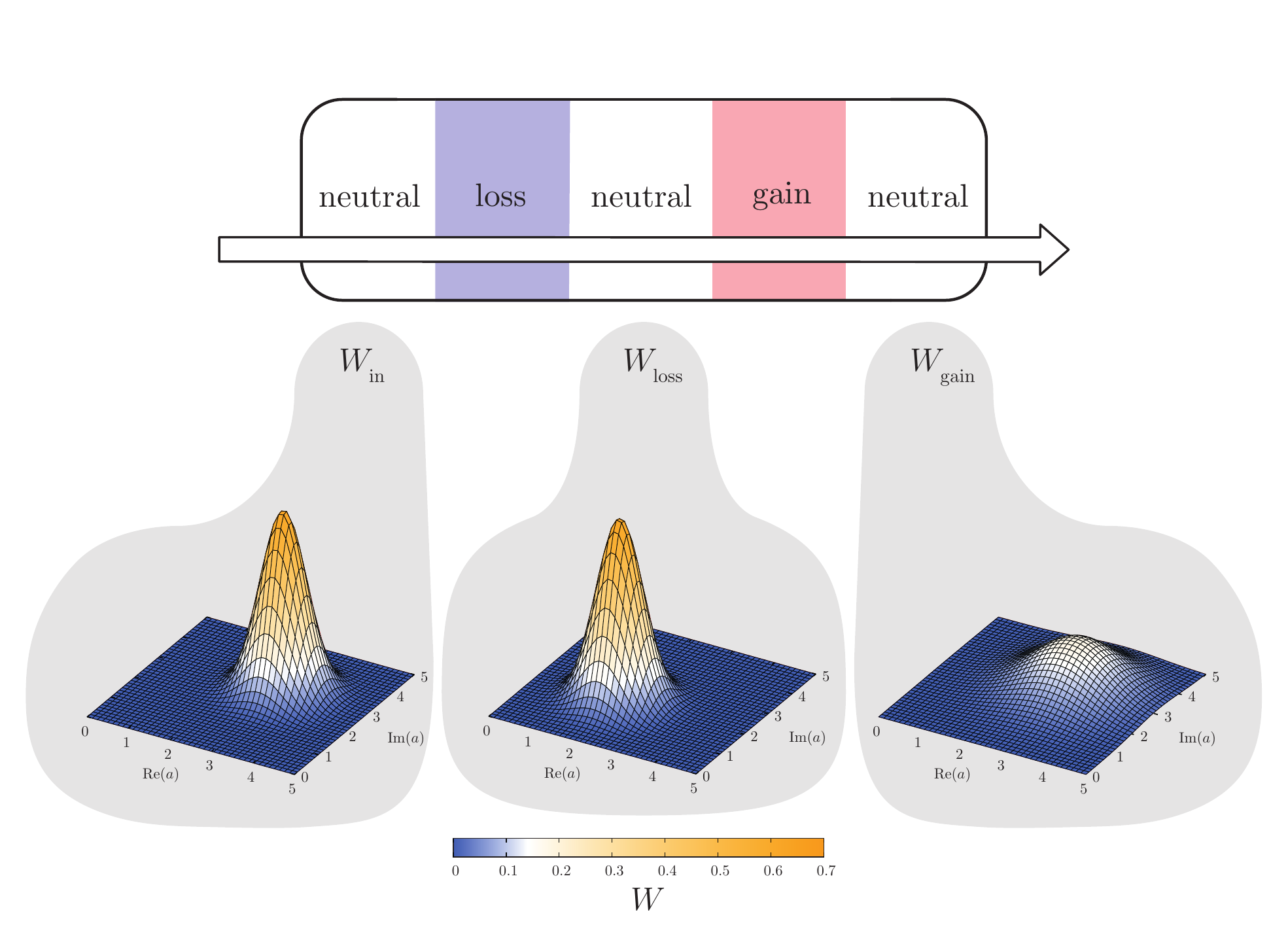}
\caption{Evolution of a coherent state through a concatenated system of loss and gain. Shown on the left
is the initial Wigner function of a coherent state with $a_0=3+3i$, in the center after transmission through an optical device with transmission coefficient
$T=2/3$, and on the right after propagation through a gain medium with $G=1/T$. It can be clearly observed that the state
significantly changes. \label{fig:wignerfunctions}}
\end{figure}

We illustrate this fundamental difference by the propagation of a coherent
quantum state through a system that consists of concatenated regions of loss
and gain. In Fig.~\ref{fig:wignerfunctions} we show the Wigner function (a
phase-space distribution function that is formally equivalent to the quantum
state \cite{Glauber69}) of a coherent quantum state with coherent amplitude
$a_0=3+3i$ (left), after transmission through an optical device with
transmission coefficient $T=2/3$ (center), and after propagation through a gain
medium with $G=1/T$ (right). One clearly observes that initial and final states
after propagation through media with loss and subsequent gain are not
equivalent.

Despite the apparent simplicity of our example, it does have far-reaching
consequences for any attempt to study $\PT$-symmetry of quantum optical
systems. First, gain always adds thermal noise to a quantum state, no matter
how gain and loss are spatially distributed. Second, we have chosen coherent
quantum states that are, on the one hand, minimum-uncertainty states that
closely resemble classical states \cite{Glauber69} and, on the other hand, have
the unique property that their purity is not affected by loss. Any other type
of quantum state will already be drastically altered by an absorbing medium.
For example, pure photon-number states turn into mixed states with lower photon
numbers \cite{Knoll99}. As a consequence, any quantum state of light is
crucially altered when propagating through any distribution of gain and loss.
In other words, in a concatenated gain/loss system the quantum eigenstates
of a Hamiltonian can never be eigenstates of the $\PT$-operator, such that
for quantum states there is always $\lbrack \hat{H},\hat{\PT} \rbrack \not = 0$.
This brings us to the conclusion that in such systems $\PT$-symmetric quantum
optics does not exist.

\begin{figure}[h!]
\includegraphics[width=\columnwidth]{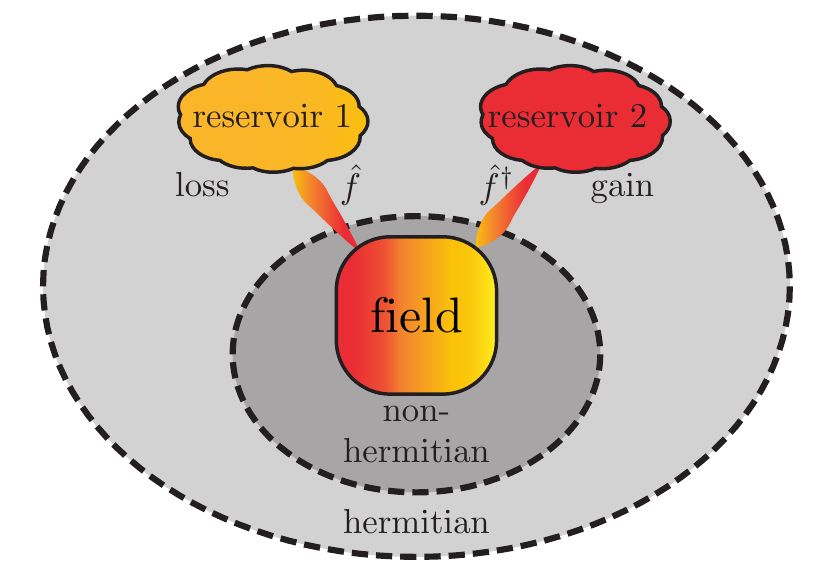}
\caption{Photonic quantum systems with gain and loss. In order to describe a non-Hermitian quantum system with gain and loss, it has to be coupled to external reservoirs that act as source and sink, respectively. This coupling introduces Langevin noise $\hat{f}$ and $\hat{f}^\dagger$ to ensure the Hermiticity of the full scheme. This noise, however, always alters propagating quantum states, even the eigenstates of the Hamiltonian of the quantum system. \label{fig:sketch}}
\end{figure}

The reason behind this surprising result is the fact that, in order to obtain
amplification, the quantum system under study has to be coupled to an external
reservoir that provides the necessary energy input. This coupling
necessarily introduces noise that can be cast into a form similar to
Eq.~(\ref{eq:langevin}),
\begin{equation}
 \dot{\hat{a}} = (-i\omega+\Gamma) \hat{a} +\hat{f}^\dagger\,,
\end{equation}
where the (Langevin) noise operators fulfil the same commutation relations as
before. The crucial difference between loss and gain is the way in which the
external reservoir is coupled to the quantum system (see Fig. \ref{fig:sketch}). In case of gain, the
noncompact SU(2,2) group transformation implies a build-up of quantum
correlations between system and reservoir that, when only observing the system,
are destroyed and manifest themselves as thermal fluctuations. This is a
similar mechanism as that observed in two-mode squeezing (which is described
by a SU(1,1) transformation) where the quantum correlations of the two squeezed
modes result in thermal distributions of the individual modes. Therefore, there
is no quantum gain mechanism that can compensate for any quantum loss process.

Importantly, our result does not only hold for harmonic oscillator modes such
as photons. Indeed, any system in which a complex potential is derived from a
coupling to a reservoir and, accordingly, admits a description by a Langevin
equation suffers from a similar conclusion. Hence, our conclusions are not
restricted to bosonic systems or, indeed, photons, but also hold for fermionic
systems, such as electrons.

To summarize our work, we have shown that $\PT$-symmetric
quantum optics is not possible within the current perception of
implementing $\PT$-symmetry, that is, using gain and loss. However, we foresee three alternative approaches
that might allow the observation of phenomena arising from $\PT$-symmetry in the
quantum realm. First, one would have to realize a complex $\PT$-invariant Hamiltonian
without coupling the quantum system to an external reservoir. However, to date such a
concept is elusive. Second, one would need to construct noiseless amplifiers. Yet,
for deterministic gain processes this violates the No-cloning theorem of quantum
mechanics \cite{Wootters82}. Indeed, relaxing the determinism constraint and allowing
probabilistic processes may result in devising probabilistic noiseless amplifiers
\cite{Xiang10}. An altogether different route
involves replacing the active gain by yet another passive loss medium such that the
overall system is lossy, which results in so-called passive $\PT$-symmetric systems.
Such structures are already implemented \cite{Guo09, Eichelkraut13, Weimann17} and seem
promising candidates for observing physics akin to $\PT$-symmetric quantum optics.

The authors thank Mark Kremer for helping to prepare the figures and acknowledge the Deutsche
Forschungsgemeinschaft (grant BL 574/13-1) for financial support.

\appendix*
\section{Appendix}
The idea of our approach is to discretize the time evolution of the electromagnetic field and
construct input-output relations between photonic amplitude operators before
and after the propagation through an optical device \cite{Scheel00}. Let the
amplitude operators of the radiation field at frequency $\omega$ at the input
of the optical device be denoted by $\hat{\mathbf{a}}$, the corresponding
output amplitude operators by $\hat{\mathbf{b}}$, and the (Langevin) operators
associated with the device by $\hat{\mathbf{g}}$. Then, the following
input-output relations read
$ \hat{\mathbf{b}} = \mathbf{T}\hat{\mathbf{a}} +\mathbf{A}\hat{\mathbf{d}}$
where the transformation and absorption matrices satisfy
 $\mathbf{T}\mathbf{T}^+ +\sigma \mathbf{A}\mathbf{A}^+ = \mathbf{I}$
and $\sigma=+1$, $\hat{\mathbf{d}}=\hat{\mathbf{g}}$ for absorption and
$\sigma=-1$, $\hat{\mathbf{d}}=\hat{\mathbf{g}}^\dagger$ for amplification.

Although a unitary evolution of the field operators alone is no longer
possible, one can nevertheless construct a unitary evolution of the combined
field-device system. Define the four-vector operators
$\hat{\bm{\alpha}}=(\hat{\mathbf{a}},\hat{\mathbf{d}})^T$ and
$\hat{\bm{\beta}}=(\hat{\mathbf{b}},\hat{\mathbf{f}})^T$ where
$\hat{\mathbf{f}}=\hat{\mathbf{h}}$ for absorption and
$\hat{\mathbf{f}}=\hat{\mathbf{h}}^\dagger$ for amplification with some
auxiliary bosonic device variables $\hat{\mathbf{h}}$. Then, the input-output
relations can be elevated to a unitary relation between the four-vector
operators as
 $\hat{\bm{\beta}}= \bm{\Lambda} \hat{\bm{\alpha}}$
with
\begin{equation*}
 \bm{\Lambda} \mathbf{J} \bm{\Lambda}^+ = \mathbf{J}\,,\quad
 \mathbf{J}=\begin{pmatrix} \mathbf{I}&\mathbf{0}\\
\mathbf{0}&\sigma\mathbf{I}\end{pmatrix}\,.
\end{equation*}
If one introduces the commuting positive Hermitian matrices
$\bm{C}=\sqrt{\mathbf{TT}^+}$ and $\bm{S}=\sqrt{\mathbf{AA}^+}$, then the
unitary matrix $\bm{\Lambda}$ can be written as \cite{Scheel00}
\begin{equation*}
 \bm{\Lambda} =
 \begin{pmatrix}
  \mathbf{T} & \mathbf{A} \\ -\sigma\mathbf{SC}^{-1}\mathbf{T} &
\mathbf{CS}^{-1}\mathbf{A}
 \end{pmatrix}\,.
\end{equation*}

The input-output relation for the amplitude operators can be cast into a
quantum-state transformation formula. Let the density operator of the input
quantum state be given as a functional of the amplitude operators
$\hat{\bm{\alpha}}$ and $\hat{\bm{\alpha}}^\dagger$, $\hat{\rho}_\textrm{in}=$
$\hat{\rho}_\textrm{in}[\hat{\bm{\alpha}},\hat{\bm{\alpha}}^\dagger]$, then the
transformed quantum state at the output is
 $\hat{\rho}_\textrm{out}=
\hat{\rho}_\textrm{in}\left[\mathbf{J}\bm{\Lambda}^+\mathbf{J}\hat{\bm{\alpha}},
\mathbf{J}\bm{\Lambda}^T\mathbf{J}\hat{\bm{\alpha}}^\dagger \right] $.
Taking the partial trace over the device variables laeves one with the quantum
state of the radiation field alone.

The equivalence between density operators and quasi-probability functions
implies a similar transformation rule for the phase-space functions. However,
care needs to be taken as the SU(2,2) transformation associated with gain mixes
creation and annihilation operators, except for the Wigner function associated
with symmetric operator ordering for which
 $W_\mathrm{out}(\bm{\alpha}) = W_\mathrm{in}
\left(\mathbf{J}\bm{\Lambda}^+\mathbf{J}\bm{\alpha}\right)$
holds.

The above relations can now be used to construct quantum states after
propagation through lossy and amplifying media. The simplest example is a
coherent state $|a_0\rangle$ whose Wigner function is given by the Gaussian
 $W(a) = \frac{2}{\pi} \exp\left(-2|a-a_0|^2\right)$
with obvious generalization for multimode states. At lossy devices, a two-mode
coherent state $|\mathbf{a}\rangle$ results in a Wigner function
 $W_\mathrm{out}(\mathbf{a}) = \int d^2g\,W_\mathrm{out}(\bm{\alpha})
 = \left( \frac{2}{\pi}\right)^2
\exp\left(-2|\mathbf{a}-\mathbf{Ta}_0-\mathbf{Ag}_0|^2\right)$
which again represents a coherent state $|\mathbf{Ta}_0+\mathbf{Ag}_0\rangle$.
If we take the device to be initially in its vacuum state,
$\mathbf{g}_0=\mathbf{0}$, we are left with a coherent
state $|\mathbf{Ta}_0\rangle$.

In the case of gain, a lengthy but straightforward application of the
quantum-state transformation relations shows that the same coherent state
transforms into
\begin{eqnarray*}
W_\mathrm{out}(\mathbf{a}) = \left(\frac{2}{\pi}\right)^2
\frac{1}{\det(2\mathbf{TT}^+-\mathbf{I})} \times \\
\times \exp\left[-2\left(\mathbf{a}^+-\mathbf{a}_0^+\mathbf{T}^+\right)
(2\mathbf{TT}^+-\mathbf{I})^{-1}
\left(\mathbf{a}-\mathbf{T}\mathbf{a}_0\right) \right]
\end{eqnarray*}
which is no longer a coherent state, but a displaced thermal state whose
temperature depends on the gain. Neglecting reflection at the interface of the
device, the (single-mode) Wigner function of the transmitted light is
\begin{equation*}
 W_\mathrm{out}(a) = \frac{2}{\pi} \frac{1}{2|T|^2-1}
\exp\left[-\frac{2|a-Ta_0|^2}{2|T|^2-1}\right]
\end{equation*}
where the transmission coefficient $|T|>1$ due to gain.

If we now construct a hypothetical device that consists of a sequence of a
lossy medium with transmission coefficient $|T|<1$ followed by a gain medium
with transmission coefficient $|G|=1/|T|>1$, this would mimic a PT-symmetric
system. However, as we have seen, noise enters both during the absorption as
well as the amplification process. In fact, starting with a (single-mode)
Wigner function
 $W_\mathrm{in}(a) = \frac{2}{\pi} \exp\left(-2|a-a_0|^2\right) $,
after propagation through a lossy medium this turns into
 $W_\mathrm{loss}(a) = \frac{2}{\pi} \exp\left(-2|a-Ta_0|^2\right) $.
Reversing the loss by amplification then results in a Wigner function
\begin{equation*}
 W_\mathrm{gain}(a) = \frac{2}{\pi} \frac{1}{2|G|^2-1}
\exp\left[-\frac{2|a-a_0|^2}{2|G|^2-1}\right]
\end{equation*}
which is a (thermally) broadened version of the original Wigner function with
the mean thermal photon number $n_\mathrm{th} = |G|^2-1$ or, equivalently,
$T_\textrm{eff}=-\frac{\hbar\omega}{k_B}\ln(1-|T|^2)$.
What it also shows is that only the first-order moments of the amplitude
operators are conserved by this system, not even the second-order moments.
Indeed, the mean number of photons contained in a quantum state with Wigner
function $W_\mathrm{gain}(a)$ is
 $\langle\hat{n}\rangle = \int d^2a\, \left( |a|^2 -\frac{1}{2} \right)
W_\mathrm{gain}(a) = |a_0|^2 + n_\textrm{th}$,
which deviates from the coherent state result by the addition of the mean
thermal photon number associated with the gain.

\end{document}